\documentclass[fleqn,usenatbib]{mnras}

\usepackage{newtxtext,newtxmath}
\usepackage{lscape}
\usepackage{pdflscape}

\usepackage[T1]{fontenc}
\DeclareRobustCommand{\VAN}[3]{#2}
\let\VANthebibliography\thebibliography
\def\thebibliography{\DeclareRobustCommand{\VAN}[3]{##3}\VANthebibliography}

\usepackage{graphicx}	% Including figure files
\usepackage{amsmath}	% Advanced maths commands
\usepackage{booktabs,longtable}
\usepackage{xtab,afterpage}
\usepackage{adjustbox}
\extrafloats{200}
\usepackage{caption}
\usepackage{soul}
\usepackage{xcolor}

\usepackage{subcaption}
\title[MPTA: First search for gravitational waves]{The MeerKAT Pulsar Timing Array: The first search for gravitational waves with the MeerKAT radio telescope}

\author[Miles et al.]{
Matthew~T.~Miles$^{1,2}$\thanks{E-mail: \href{mailto:matthewmiles@swin.edu.au}{matthewmiles@swin.edu.au}},
Ryan~M.~Shannon$^{1,2}$,
Daniel~J.~Reardon$^{1,2}$,
Matthew~Bailes$^{1,2}$,
David~J.~Champion$^{3}$,\newauthor
Marisa~Geyer$^{4}$,
Pratyasha~Gitika$^{1,2}$,
Kathrin~Grunthal$^{3}$,
Michael~J.~Keith$^{5}$,
Michael~Kramer$^{3,5}$,\newauthor
Atharva~D.~Kulkarni$^{1,2}$,
Rowina~S.~Nathan$^{6,2}$,
Aditya~Parthasarathy$^{7,8,3}$,
Jaikhomba~Singha$^{4}$,
Gilles~Theureau$^{9,10}$,\newauthor
Eric~Thrane$^{6,2}$,
Federico~Abbate$^{11,3}$,
Sarah~Buchner$^{12}$,
Andrew~D.~Cameron$^{1,2}$,
Fernando~Camilo$^{12}$,\newauthor
Beatrice~E.~Moreschi$^{13,11}$,
Golam~Shaifullah$^{13,11}$,
Mohsen~Shamohammadi$^{1,2}$,
Andrea~Possenti$^{11}$ \& \newauthor
Vivek~Venkatraman~Krishnan$^{3}$
\\
$^{1}$Centre for Astrophysics and Supercomputing, Swinburne University of Technology, PO Box 218, Hawthorn, VIC 3122, Australia\\
$^{2}$OzGrav: The ARC Centre of Excellence for Gravitational Wave Discovery \\
$^{3}$MPI f\"ur Radioastronomie, Auf dem H\"ugel 69, 53121 Bonn, Germany \\
$^{4}$High Energy Physics, Cosmology \& Astrophysics Theory (HEPCAT) Group, Department of Mathematics and Applied Mathematics, \\ University of Cape Town, Cape Town 7700, South Africa \\
$^{5}$Jodrell Bank Centre for Astrophysics, University of Manchester, Alan-Turing Building, Oxford Street, Manchester M13 9PL, UK \\
$^{6}$School of Physics and Astronomy, Monash University, Clayton VIC 3800, Australia \\
$^{7}$ASTRON, Netherlands Institute for Radio Astronomy, Oude Hoogeveensedijk 4, 7991 PD Dwingeloo, The Netherlands \\
$^{8}$Anton Pannekoek Institute for Astronomy, University of Amsterdam, Science Park 904, 1098 XH Amsterdam, The Netherlands \\
$^{9}$Laboratoire de Physique et Chimie de l'Environnement et de l'Espace  LPC2E UMR7328, Université d'Orléans, CNRS, CNES, OSUC,\\ Observatoire de Paris,  F-45071 Orléans, France \\
$^{10}$Laboratoire Univers et Théories, Observatoire de Paris, Université PSL, Université de Paris Cité,  CNRS, F-92190 Meudon, France \\
$^{11}$INAF -- Osservatorio Astronomico di Cagliari, Via della Scienza 5, I-09047 Selargius (CA), Italy \\
$^{12}$South African Radio Astronomy Observatory, 2 Fir Street, Black River Park, Observatory 7925, South Africa \\
$^{13}$Dipartimento di Fisica ``G. Occhialini", Universit{\'a} degli Studi di Milano-Bicocca, Piazza della Scienza 3, I-20126 Milano, Italy\\
}

\date{Accepted XXX. Received YYY; in original form ZZZ}

\pubyear{2024}

\begin{document}
\label{firstpage}
\pagerange{\pageref{firstpage}--\pageref{lastpage}}
\maketitle

\begin{abstract}
\newline

Pulsar Timing Arrays search for nanohertz-frequency gravitational waves by regularly observing ensembles of millisecond pulsars over many years to look for correlated timing residuals. Recently the first evidence for a stochastic gravitational wave background has been presented by the major Arrays, with varying levels of significance ($\sim$2-4$\sigma$). In this paper we present the results of background searches with the MeerKAT Pulsar Timing Array. Although of limited duration (4.5 yr), the $\sim$ 250,000 arrival times with a median error of just $3 \mu$s on 83 pulsars make it very sensitive to spatial correlations. Detection of a gravitational wave background requires careful modelling of noise processes to ensure that any correlations represent a fit to the underlying background and not other misspecified processes. Under different assumptions about noise processes we can produce either what appear to be compelling Hellings-Downs correlations of high significance (3-3.4$\sigma$) with a spectrum close to that which is predicted, or surprisingly, under slightly different assumptions, ones that are insignificant. This appears to be related to the fact that many of the highest precision MeerKAT Pulsar Timing Array pulsars are in close proximity and dominate the detection statistics. The sky-averaged characteristic strain amplitude of the correlated signal in our most significant model is $h_{c, {\rm yr}} = 7.5^{+0.8}_{-0.9} \times 10^{-15}$ measured at a spectral index of $\alpha=-0.26$, decreasing to $h_{c, {\rm yr}} = 4.8^{+0.8}_{-0.9} \times 10^{-15}$ when assessed at the predicted $\alpha=-2/3$. These data will be valuable as the International Pulsar Timing Array project explores the significance of gravitational wave detections and their dependence on the assumed noise models.

\end{abstract}


\section{Introduction}

The first direct detection of gravitational waves (GWs) \citep{2016PhRvL.116f1102A} marked a rare occurrence in physics and astronomy: the beginning of a new field, and a new way to study the Universe. Almost a century passed between the initial prediction of GWs by Einstein \citep{1916AnP...354..769E} and the observation of perturbed space-time from the coalescence of two black holes thirty times more massive than the Sun. During this time the very existence of GWs had been called into question (at one stage by Einstein himself), and early attempts to detect them had been unsuccessful \citep{2016Univ....2...22C}. The Hulse-Taylor binary pulsar \citep{1975ApJ...195L..51H} had provided compelling indirect evidence
that gravitational waves were emitted at the rate implied by the General theory of relativity \citep{1982ApJ...253..908T}.  
Decades later, there now exists a catalogue of GW detections that continues to grow \citep{2021arXiv211103606T} for observations limited to the \textit{audio band}, corresponding to gravitational waves that can be observed at frequencies between $\sim 10$ to $\sim 1000$ Hz. Pulsar timing arrays (PTAs) offer an alternate window into the GW-bright Universe, sensitive to GWs at nanohertz (nHz) frequencies. Potential sources of these include inspiralling supermassive black hole binaries (SMBHBs) \citep{1995ApJ...446..543R, 2003ApJ...583..616J, 2003ApJ...590..691W, 2004ApJ...611..623S, 2011MNRAS.411.1467K, 2012A&A...545A.127R,2012ApJ...761...84R, 2017PhRvL.118r1102T}, cosmic strings \citep{1976JPhA....9.1387K, PhysRevD.81.104028, 2012PhRvD..85l2003S, 2015MNRAS.453.2576L, 2018ApJ...859...47A}, cosmological phase transitions, \citep{1980PhLB...91...99S, 2005PhyU...48.1235G} and quantum fluctuations in the early Universe \citep{2000gr.qc.....8027M, 2016PhRvX...6a1035L}. Of these, gravitational waves from SMBHBs possess the strongest theoretical motivation. Thus, the strongest most probable signal is expected to be the incoherent superposition of all gravitationally radiating, inspiralling SMBHBs in the observable Universe, referred to as the stochastic gravitational wave background (SGWB). 

PTAs are regularly observed ensembles of millisecond pulsars (MSPs) in which the arrival times of pulses emitted by pulsars are measured over years to decades \citep{1990ApJ...361..300F}. Of all pulsars, MSPs are the most rotationally stable, allowing the times of arrival (TOAs) of their pulses to be predicted to a sub-microsecond precision in some systems. The predictability of their emission allows for a wide range of insights, from the properties of pulsar local environments \citep{Wolszczan_Frail_1992, Wolszczan_planets_1994}, the nature of the nuclear equation of state \citep{Demorest_Nature_2010, Antoniadis1233232, 2021ApJ...915L..12F, 2021arXiv210506979M, 2021arXiv210506980R}, to stringent tests of gravity \citep{2003LRR.....6....5S, 2021PhRvX..11d1050K}. It also enables an ensemble of pulsars (a PTA) to be searched for common signals such as the influence of an SGWB. This signal is thought to emerge in PTA data sets both as a noise process with consistent spectral properties in individual pulsar arrival times (the auto-correlated signal) and as a temporally and an angularly correlated signal between pulsars that is a function of their angular separation (the cross-correlated signal). The auto-correlated signal is expected to emerge first in the data of most pulsar timing arrays \citep{2013CQGra..30v4015S}, and a signal that is representative of this has been observed in multiple PTAs \citep{NanoGravGWB, 2021arXiv210712112G, 2021MNRAS.508.4970C, 2022MNRAS.510.4873A, 2023RAA....23g5024X}. However, it has also been shown that such a signal can arise spuriously \citep{2022MNRAS.516..410Z, 2022ApJ...932L..22G}. As such, it is only a possible indicator that an SGWB signal may be present in the data, rather than evidence for it. Instead, the signature that is sought after is the cross-correlated signal, which shows angular correlations that are the consequence of the quadrupolar nature of gravitational waves, distinct from other correlated signals in PTA data sets \citep{2016MNRAS.455.4339T}. The detection of an SGWB is the primary focus of PTA collaborations, with searches having been performed by the European PTA \citep[EPTA;][]{2008AIPC..983..633J}, the Parkes PTA \citep[PPTA;][]{2013PASA...30...17M}, the North American Nanohertz Observatory for Gravitational Waves \citep[NANOGrav;][]{2009arXiv0909.1058J}, the Chinese PTA \citep[CPTA;][]{2023RAA....23g5024X}, and also in a joint effort through the International PTA \citep[IPTA;][]{2010CQGra..27h4013H}. The searches have not yet exceeded IPTA-nominated thresholds to claim a detection \citep{2023arXiv230404767A}. However, strong evidence ($3-4\sigma$) has recently been shown for an angularly correlated signal \citep{2023RAA....23g5024X, 2023arXiv230616214A, 2023ApJ...951L...8A, 2023ApJ...951L...6R}, suggesting a definitive detection could be imminent. 
The primary target of GW searches with PTAs is an isotropic background. However, it is possible that the background may emerge as an anisotropic signal from the influence of bright SMBHBs \citep{2013PhRvD..88h4001T, 2017NatAs...1..886M}. In the era of an initial detection of an SGWB, whether it is anisotropic may not be immediately obvious, and identifying this is a goal that could benefit from the combination of data from multiple PTAs by the IPTA. Individual inspiralling SMBHBs can also emit GWs in the nHz-frequency band that can potentially be detected by PTAs. These continuous gravitational waves are expected to initially appear in PTA data as non-evolving signals, confined to a single frequency characteristic of their emission. Their evolution occurs over a timescale much longer than what current PTAs can observe, making them detectable as static signals in PTA data. While an SGWB is thought to be be the strongest signal in PTA data, bright individual binaries may also be detectable in tandem with, or soon after, a confirmed detection of an SGWB \citep{2009MNRAS.394.2255S, 2015MNRAS.451.2417R}.

In this paper, we describe the first searches for gravitational waves with the MeerKAT Pulsar Timing Array (MPTA), reporting evidence for nHz-frequency gravitational waves in a $4.5$-year MPTA data set that appears dependent on the assumptions that we make. In Section \ref{ch6: methodology} we describe the methodology, data set, and noise models used for this analysis. The results of the searches are presented in Section \ref{ch6: results}. The implications of these are discussed in Section \ref{ch6: discussion}, and we state our conclusions in Section \ref{ch6: conclusions}.

\section{Methodology}
\label{ch6: methodology}   

The data set that is used for this analysis is described in \citet{MPTA2024_data+noise} that extends on from \citet{2023MNRAS.519.3976M}, comprising $4.5$ years of observations recorded with the MeerKAT L-band receiver ($856 - 1712$ MHz).

\subsection{Search for the correlated signature of an SGWB}

To determine the presence of all signals in the MPTA data, we use the standard PTA likelihood
\begin{equation}
\label{eq: ch6 PTA likelihood}
    \mathcal{L}(d | \theta) = \frac{\mathrm{exp}(-\frac{1}{2}\mathbf{\delta t}^{T}\mathbf{C}^{-1}\mathbf{\delta t})}{\sqrt{\mathrm{det}(2\pi \mathbf{C})}},
\end{equation}
where $\mathbf{\delta t}$ is a vector of timing residuals, $\mathbf{C}$ is the $N_\mathrm{TOA} \times N_\mathrm{TOA}$ covariance matrix of the data, and $\theta$ describes the parameters of the model \citep{2009MNRAS.395.1005V}. 

We assume that the gravitational-wave is described by a power-law so that its power spectral density (in the residuals) is of the form
\begin{equation}
    \mathrm{P}(f; \mathrm{A}, \gamma) = \frac{\mathrm{A}^{2}}{12 \pi^{2}}\left(\frac{f}{f_\mathrm{c}}\right)^{-\gamma} \mathrm{yr}^{3},
\end{equation}
where $A$ is the characteristic strain amplitude of the signal referenced to a frequency of $f_\mathrm{c}$, $f$ is gravitational-wave frequency,  and $\gamma$ is the spectral index of the process. We  set the reference frequency to be $1\,\mathrm{yr}^{-1}$, as is standard in PTA analysis. 

The angularly correlated component of this signal must also be modelled, which for an isotropic background is defined by the Hellings-Downs correlation function \citep{Hellings_Downs_1983}:
\begin{equation}
\label{eq: ch6_HD}
    \Gamma_\mathrm{a,b}(\zeta) = \frac{1}{2} - \frac{1}{4}\left(\frac{1 - \cos{\zeta}}{2} \right) + \frac{3}{2}\left(\frac{1 - \cos{\zeta}}{2}\right)\ln\left({\frac{1 - \cos{\zeta}}{2}}\right),
\end{equation}
where $\zeta$ is the angular separation of a pair of pulsars, $a$ and $b$, within the PTA. For simplicity, we define the signal expected from an SGWB as the product of the two, such that
\begin{equation}
\label{eq: HD_comb_signal}
    \mathrm{P_{HD}} = \mathrm{P}(f; \mathrm{A}, \gamma)\Gamma_\mathrm{a,b}(\zeta).
\end{equation}

We assess the presence of this signal in the data through Bayesian, frequentist, and \textit{pseudo-Bayesian} frameworks. In the frequentist framework, we make use of the optimal statistic (OS) \citep{2009PhRvD..79h4030A, 2015PhRvD..91d4048C}. While doing this, we employ noise marginalisation using the output of a Bayesian search for an uncorrelated common signal using comprehensive noise models. In \citet{MPTA2024_data+noise}, a free spectrum analysis of the common signal identified in the data is assessed, finding that the amplitude measurement is constrained only in the three lowest independent frequencies. Due to this, we undertake the frequentist analyses using only these frequencies, the lowest of which corresponds to $1/T$, where $T$ is the observing span of this MPTA data set. Between $1/T$ and $4/T$, where the PTA is no longer sensitive, these harmonics correspond to a range of frequencies between $7.1$ and $28.4$ nHz.

We assess the significance of the frequentist analysis by employing bootstrap methods that are commonly used in PTA analyses, known as sky and phase scrambling \citep{2016PhRvD..93j4047C, 2017PhRvD..95d2002T}. In the Bayesian framework, we employ the Savage-Dickey method \citep{10.1214/aoms/1177693507} to compute a Bayes factor ($\mathcal{B}$) to select the model that is preferred by the data and assess its significance. To complement these analysis streams, we also provide an assessment that acts as an intermediate between both, which we refer to as a \textit{pseudo-Bayesian} pairwise analysis \citep{2023ApJ...951L...6R}.

\subsection{Search for a single source}
\label{subsection: ch6_single_source_methods}

In addition to searching for correlated signals emerging from an SGWB, we also search over the MPTA data set for the presence of a single gravitational wave source that is emitting a non-evolving continuous gravitational wave (CW). We construct this model assuming an individual SMBHB, emitting a CW at a single frequency. We search for a circular binary both in the presence and absence of an SGWB. As only the three lowest frequency bins are constrained when assessing the common uncorrelated signal via a free spectrum analysis, if a CW is present in the MPTA data, it is possible that spectral leakage could result in the signal being spread between these frequencies. This might result in difficulties detecting the presence of both an SGWB and CW signal if it existed in the most sensitive frequencies of the array; however, for consistency with analyses performed by other PTAs, we search for both simultaneously.

To model the CW signal, we follow the approaches taken in \citet{2023ApJ...951L..28A} and \citet{2023ApJ...951L..50A}, which we summarise here. The signal of a single GW emitting source can be modelled in pulsar timing data to be
\begin{equation}
    \label{eq: single_source_signal}
    \mathrm{s(t)} = F^{+}(\theta, \phi, \psi) [\mathrm{s}_{+}(\mathrm{t_{p}}) - \mathrm{s}_{+}(\mathrm{t})] + F^{\times}(\theta, \phi, \psi) [\mathrm{s}_{\times}(\mathrm{t_{p}}) - \mathrm{s}_{\times}(\mathrm{t})].
\end{equation}
The two terms, $\mathrm{s_{+, \times}(t)}$ and $\mathrm{s_{+, \times}(t_{p})}$, are the Earth and pulsar terms of the signal, with $+$ and $\times$ representing \textit{plus} and \textit{cross} polarisation modes of GW emission from General Relativity. 
The $F$ factors in Equation \ref{eq: single_source_signal} are the antenna pattern function of the pulsar response, which is dependent on the location of the single source ($\theta$, $\phi$) and the polarisation angle of the GW ($\psi$). 

For a circular binary, the terms are
\begin{equation}
    \mathrm{s_{+}(t)} = \frac{\mathcal{M}^{5/3}}{\mathrm{d_{L}}\omega(\mathrm{t})^{1/3}}\sin2\Phi(\mathrm{t})(1+\cos^{2}i),
\end{equation}
and
\begin{equation}
    \mathrm{s_{\times}(t)} = \frac{\mathcal{M}^{5/3}}{\mathrm{d_{L}}\omega(\mathrm{t})^{1/3}}2\cos2\Phi(\mathrm{t})\cos i,
\end{equation}
where $i$ is the inclination angle of the SMBHB, $\mathrm{d_{L}}$ is the luminosity distance, $\mathcal{M}$ is the chirp mass of the binary, $\omega(\mathrm{t})$ is the angular frequency, and $\Phi(\mathrm{t})$ is the GW phase. The angular frequency depends on the physical parameters of the binary and evolves with time, although it is not plausible for this evolution to be detectable within the length of this MPTA data set:
\begin{equation}
    \omega(\mathrm{t}) = 2\pi f_\mathrm{GW} \left[1 - \frac{256}{5}\mathcal{M}^{5/3}(2\pi f_\mathrm{GW})^{8/3}(\mathrm{t-t_{0}})\right]^{-3/8},
\end{equation}
where $f_\mathrm{GW}$ is the GW frequency at time $t_0$. A similar transformation can be performed for the phase, 
\begin{equation}
    \Phi(\mathrm{t}) = \Phi_{0} + \frac{1}{32}\mathcal{M}^{-5/3} [(2\pi f_\mathrm{GW})^{-5/3} - \omega(\mathrm{t})^{-5/3}],
\end{equation}
where $\Phi_{0}$ is the initial Earth term phase. Finally, the luminosity distance can be replaced to include an expression describing the overall amplitude of the signal in terms of characteristic strain ($\mathrm{h_{c}}$)
\begin{equation}
    \mathrm{d_{L}} = \frac{2\mathcal{M}^{5/3}(\pi f_\mathrm{GW})^{2/3}}{\mathrm{h_{c}}}.
\end{equation}

These substitutions allow a CW signal to be parameterised with only eight parameters, in conjunction with the intrinsic terms connected to the pulsars in the PTA, and the optional inclusion of a term describing an SGWB. Due to generally imprecise PTA constraints on pulsar distances, we forego the inclusion of the pulsar terms in this analysis. A more complete derivation of this signal, as well as a derivation of the more complex case of an eccentric binary, can be found in \citet{2016ApJ...817...70T}.

\subsection{Pulsar noise models}

To assess the contribution of a GW signal in the data, the comprehensive MPTA noise models described in \citet{MPTA2024_data+noise} are used in a Bayesian and frequentist methodology, consistent with other PTA analyses \citep{2023arXiv230616214A, 2023ApJ...951L...8A}, as well as a \textit{pseudo-Bayesian} method as performed by the PPTA \citep{2023ApJ...951L...6R}. We assess the presence of a common signal not only under the assumption of these noise models, but also including additional achromatic red noise in an effort to mitigate potential misspecification of sub-threshold processes that were not favoured for inclusion in \citet{MPTA2024_data+noise}. Finally, we assess the presence of the signal using a slightly modified version of this model, in which a noise process that is not well understood is removed from the set of models. This is described further below. 

We use a set of sub and super-scripts to describe the model of intrinsic pulsar noise processes that are used in this assessment: DATA to refer to the model described by the data derived noise processes defined in \citet{MPTA2024_data+noise}, ER to refer to the DATA model with additional achromatic red noise processes included, and ALT to refer to the modestly altered ER model. In the frequentist analyses of these models, we employ a solution derived of the one-dimensional probabilistic maximum for each varied parameter. In some cases, the values chosen using this method are not located in the densest regions of the joint multi-dimensional posterior distribution. Where this is the case, the values of a process are chosen from the densest regions of the multi-dimensional posterior that are coincident with the one-dimensional probabilistic maximum of the best-constrained parameter (usually the amplitude).

Hereafter, we refer to this as the FP (from \textit{fixed parameter}) solution. If the posterior of the noise process is not well constrained following an ensemble search (i.e. a Bayesian analysis of the entire PTA), the process is not included in the downstream frequentist analysis\footnote{In practice this is done by setting the amplitude of the noise process to an immeasurably small value of $\log_{10}\mathrm{A}=-20$, which is five to six orders of magnitude smaller than a signal that could be measured in the data set}.
%, it is manually removed by setting the amplitude of the signal to $\log_{10}\mathrm{A}=-20$. 
This is done to mitigate the impact of unfavoured processes on the overall assessment of a common signal. Largely, the processes that are found to be unconstrained during ensemble searches are consistent across the models we consider. In a few cases, a process would no longer be favored in only one of the assessments. As the properties of a common signal are not significantly different between the models, these cases were monitored carefully and assessed for further inclusion based on the relative support for the process between the single pulsar noise analysis and the ensemble analysis.

\begin{figure}
    \includegraphics[width=\columnwidth]{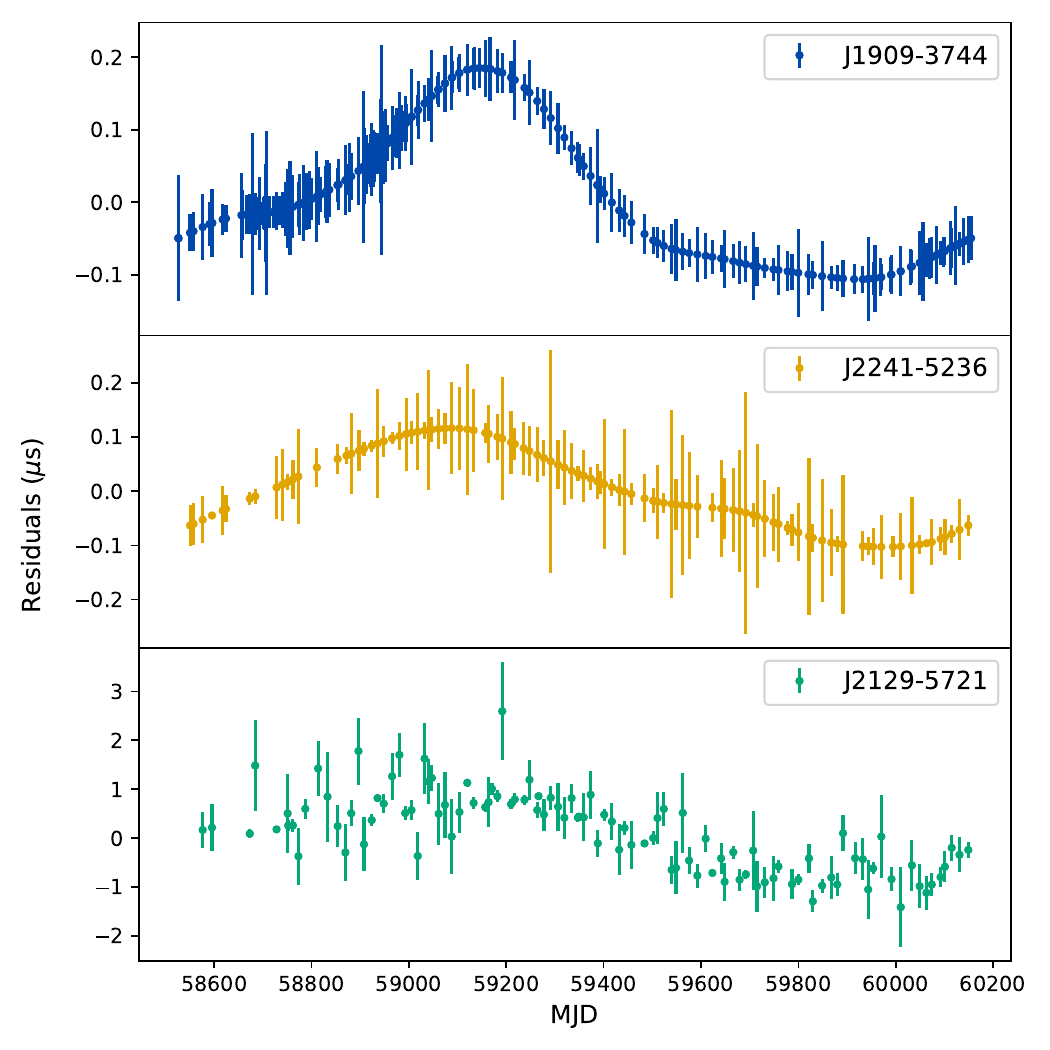}
    \caption{A comparison between time realisations of a common achromatic process for two MPTA pulsars, PSR~J1909$-$3744 (top) and PSR~J2241$-$5236 (centre), against the timing residuals of PSR~J2129$-$5721. The time realisations of the common achromatic process that we display here appear to be highly correlated, as is expected for pulsars with low angular separations under the assumption of an SGWB. In addition, the realisations appear highly correlated with the unaltered timing residuals of PSR~J2129$-$5721, which is at low angular separations with both of these pulsars.}
    \label{fig: J2129_compare}
\end{figure}

The ALT model differs from the ER model by excluding the extraordinarily steep (high spectral index) achromatic red noise process identified in PSR~J2129$-$5721. While this process was not supported for inclusion by the analysis method utilised in \citet{MPTA2024_data+noise}, its amplitude is constrained in the ensemble Bayesian assessment of the ER models. This is unexpected, as the PPTA does not report any constraints on an achromatic red noise process for this pulsar over almost two decades of observations \citep{2023ApJ...951L...7R}. Furthermore, the process is unlike any other found in the MPTA: the constraint on its spectral index extends to $\gamma > 10$, the process is only constrained in its lowest frequency bin, and the amplitude of the process in that bin is heavily degenerate with a common signal across the PTA. Due to the unusual and inconsistent nature of this noise process, we do not believe that it is well modelled using an achromatic red noise process; instead, it is likely a rare and confounding case of noise misspecification. The unusual nature of this noise is further demonstrated in Figure \ref{fig: J2129_compare}, in which the unaltered timing residuals of the pulsar are juxtaposed against time-realisations of a common achromatic process in two other pulsars that are at low angular separations to PSR~J2129$-$5721, providing circumstantial evidence that the noise processes are common between the pulsars. We provide brief descriptions of these models in Table \ref{tab: model_table}.

% \usepackage{booktabs}
\begin{table*}
\caption{Models considered in this work. We provide the number of free parameters that are used in each model and a brief description.}
\label{tab: model_table}
\begin{tabular}{@{}lcl@{}}
\toprule
Model & \multicolumn{1}{l}{Free parameters} & Model summary                                                                                                         \\ \midrule
DATA  & 390                                    & Pulsar noise processes as described in \citet{MPTA2024_data+noise} \\
ER    & 532                                    & An extension to the DATA model with additional achromatic red noise processes when not supported in DATA   \\
ALT   & 530                                    & The ER model, with the exclusion of the achromatic red noise in PSR~J2129$-$5721                                     \\ \bottomrule
\end{tabular}
\end{table*}

\section{Results}
\label{ch6: results}

We search for two correlated signals: an isotropic SGWB and a single radiating GW source. For the remainder of the paper,  we employ the subscript CURN to refer to a common signal without angular correlations, SMBHB for angular correlations representative of a single radiating source, and HD to represent an angularly correlated signal with correlations expected from an isotropic SGWB.

In \citet{MPTA2024_data+noise} we reported the presence of an uncorrelated common process in the data using the ER model, with the  characteristic strain amplitude of the power-law signal measured to be $\log_{10}\mathrm{A^{ER}_{CURN}}=-14.25^{+0.21}_{-0.36}$ and the spectral index measured to be $\gamma^\mathrm{ER}_\mathrm{CURN}=3.60^{+1.31}_{-0.89}$ at a Bayes factor of $\mathrm{ln}(\mathcal{B})=3.17$ in favour of a CURN in the data. For completeness, we also provide equivalent constraints of the data-driven model, measured at an amplitude $\log_{10}\mathrm{A^{DATA}_{CURN}}=-14.25^{+0.21}_{-0.34}$ with a spectral index of $\gamma^\mathrm{DATA}_\mathrm{CURN}=3.52^{+1.12}_{-0.90}$.

\subsection{Frequentist assessment of isotropic angular correlations}
\label{subsection: ch6_iso_freq_SGWB}

\begin{figure*}
    \includegraphics[width=0.9\linewidth]{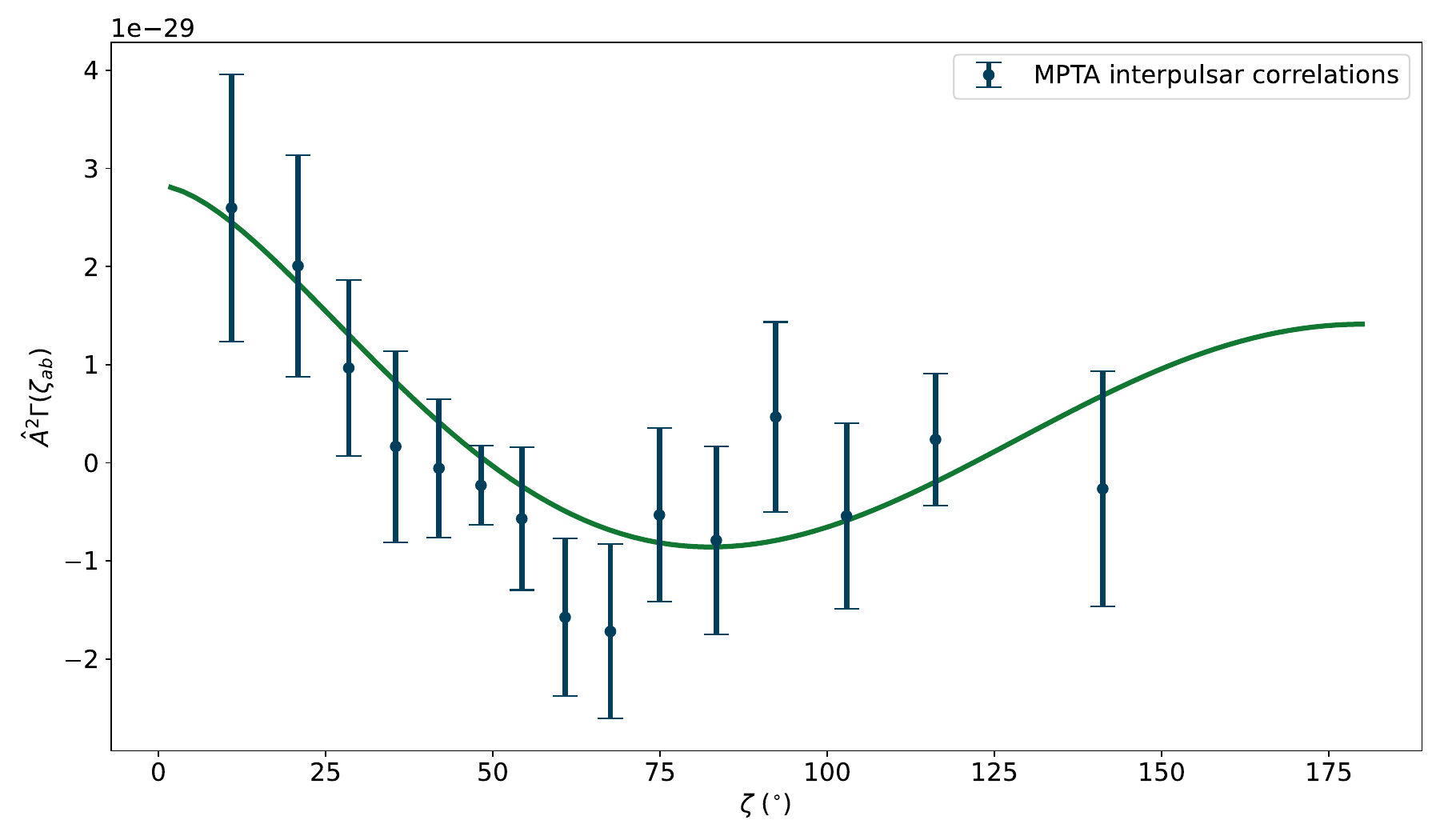}
    \caption[All-sky inter-pulsar correlations of the MPTA data set.]{All-sky inter-pulsar correlations of the MPTA data set as a function of the angular separation between pairs of pulsars, displayed in $15$ bins across the angular separation range. The $\hat{\mathrm{A}}_\mathrm{DATA}^{2} = (5.65 \pm 1.2) \times 10^{-29}$ corresponds to a dimensionless gravitational wave amplitude of $7.5^{+0.8}_{-0.9} \times 10^{-15}$.}
    \label{fig: HD_all_sky_DATA}
\end{figure*}

First, we search for this signal under our model assumptions using the frequentist optimal statistic. Using the DATA model, we measure an all-sky angularly correlated signal with an apparent signal-to-noise ratio (S/N) of $4.6$. We note that this statistic is not suitable to use as a metric of the significance of the signal, and discuss the interpretation below. For this model, this measurement is equivalent to a characteristic strain amplitude of $\hat{\mathrm{A}}_\mathrm{DATA}^{2} = (5.7 \pm 1.2) \times 10^{-29}$ ($\log_{10}\mathrm{A_{DATA}} = -14.12^{+0.04}_{-0.05}$), assuming $\gamma^\mathrm{DATA}_\mathrm{CURN}=3.52$. The presence of Hellings-Downs correlations at these values are apparent when individual pulsar pairs are averaged into bins in pulsar pair separation angle, as shown in Figure \ref{fig: HD_all_sky_DATA}, displaying the cross-correlated power in fifteen equally distributed bins for the $3403$ pulsar pairs in the MPTA. We note that both the optimal statistic amplitude and S/N are calculated on a per-pulsar-pair basis rather than from the binned averages presented in Figure \ref{fig: HD_all_sky_DATA}. The correlations that are displayed have been calculated using an algorithm described in \citet{2023PhRvD.108d3026A}, which accounts for the covariances between the pulsar pairs stemming from the similarity of their positions on the sky and the corresponding redundancy of their measurements. 

\begin{figure*}
\centering
\begin{subfigure}{0.5\linewidth}
  \centering
  \includegraphics[width=\linewidth]{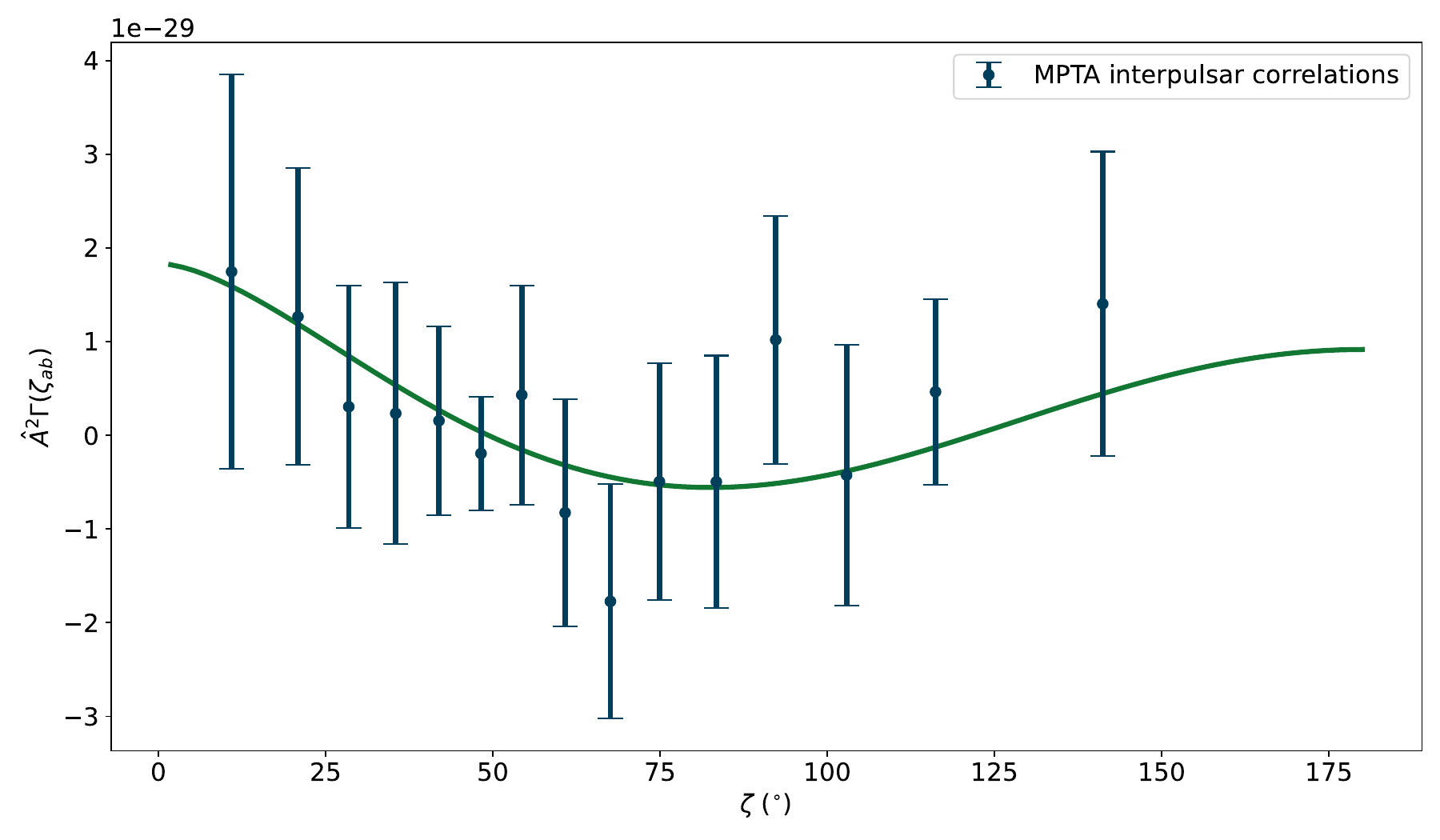}
  \caption{Angular correlations of the ER model}
  \label{fig: OS_CRN_ER}
\end{subfigure}%
\begin{subfigure}{0.5\linewidth}
  \centering
  \includegraphics[width=\linewidth]{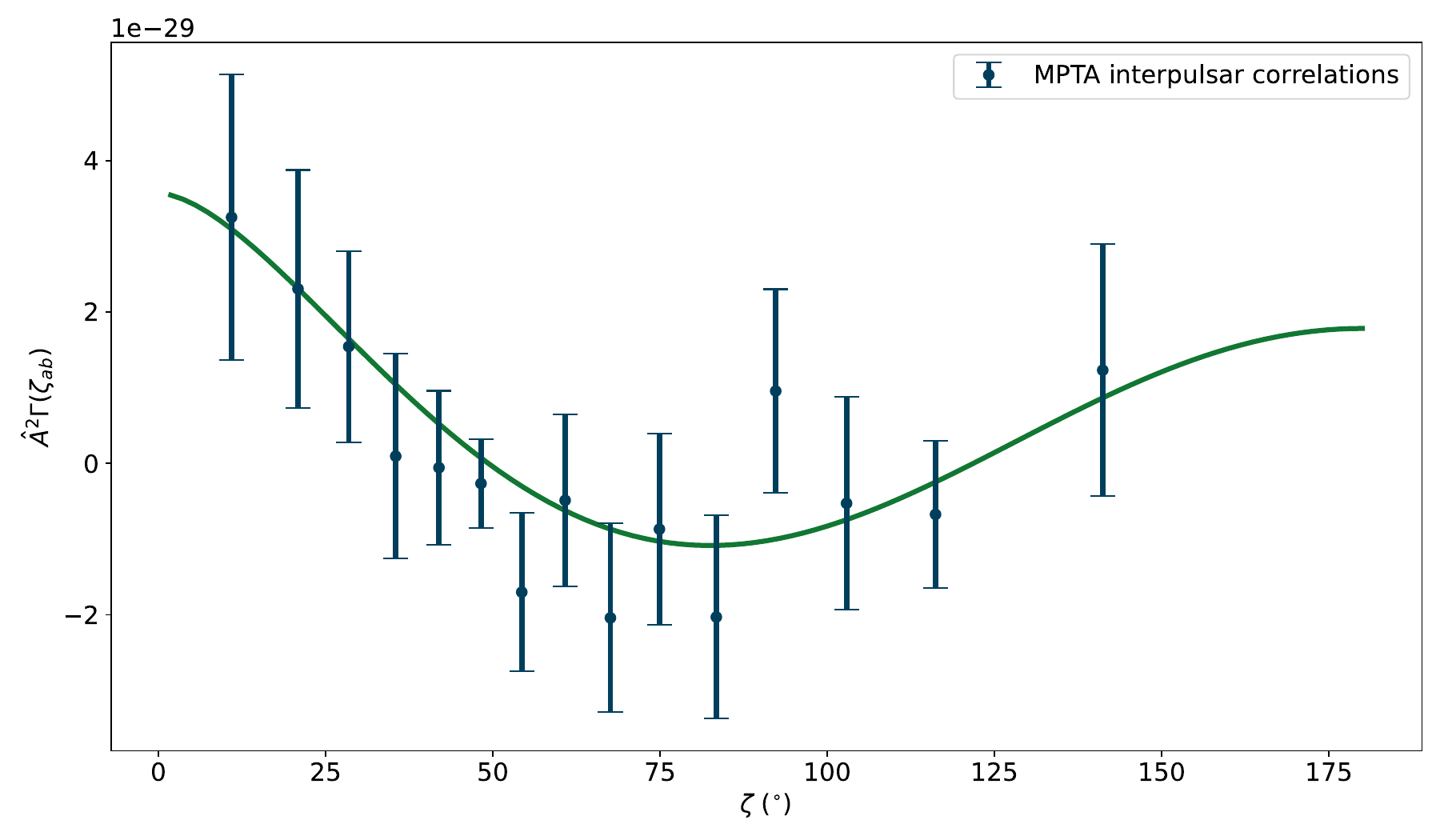}
  \caption{Angular correlations of the ALT model}
  \label{fig: OS_CRN_ALT}
\end{subfigure}
\caption{All-sky inter-pulsar correlations of the MPTA data set for the two alternative models: ER (left) and ALT (right). Using the ER model, we measure an optimal statistic amplitude to be $\hat{\mathrm{A}}_\mathrm{ER}^{2} = (3.7 \pm 1.9) \times 10^{-29}$, and for the ALT model we measure an optimal statistic amplitude to be $\hat{\mathrm{A}}_\mathrm{ALT}^{2} = (7.1 \pm 1.8) \times 10^{-29}$. While the models are very similar, there is a marked difference in the correlations that stems primarily from the lack of achromatic red noise in PSR~J2129$-$5721.}
\label{fig: OS_ER_ALT}
\end{figure*}

In an effort to be conservative in our investigation of any spatially correlated signals, we also present the correlations for the ER and ALT models. These correlations are calculated in the same manner as the DATA model, using only the marginally different assumptions in the  pulsar noise models. Assessing the ER model, we measure an optimal statistic amplitude of $\hat{\mathrm{A}}_\mathrm{ER}^{2} = (3.7 \pm 1.9) \times 10^{-29}$, the corresponding correlations of which are presented in Figure \ref{fig: OS_CRN_ER}. For the ALT model, we measure an optimal statistic amplitude of $\hat{\mathrm{A}}_\mathrm{ALT}^{2} = (7.1 \pm 1.8) \times 10^{-29}$, a statistically marginal increase in the amplitude resulting from minimal differences between the two models that can be seen in Figure \ref{fig: OS_CRN_ALT}.

\begin{figure}
    \includegraphics[width=\columnwidth]{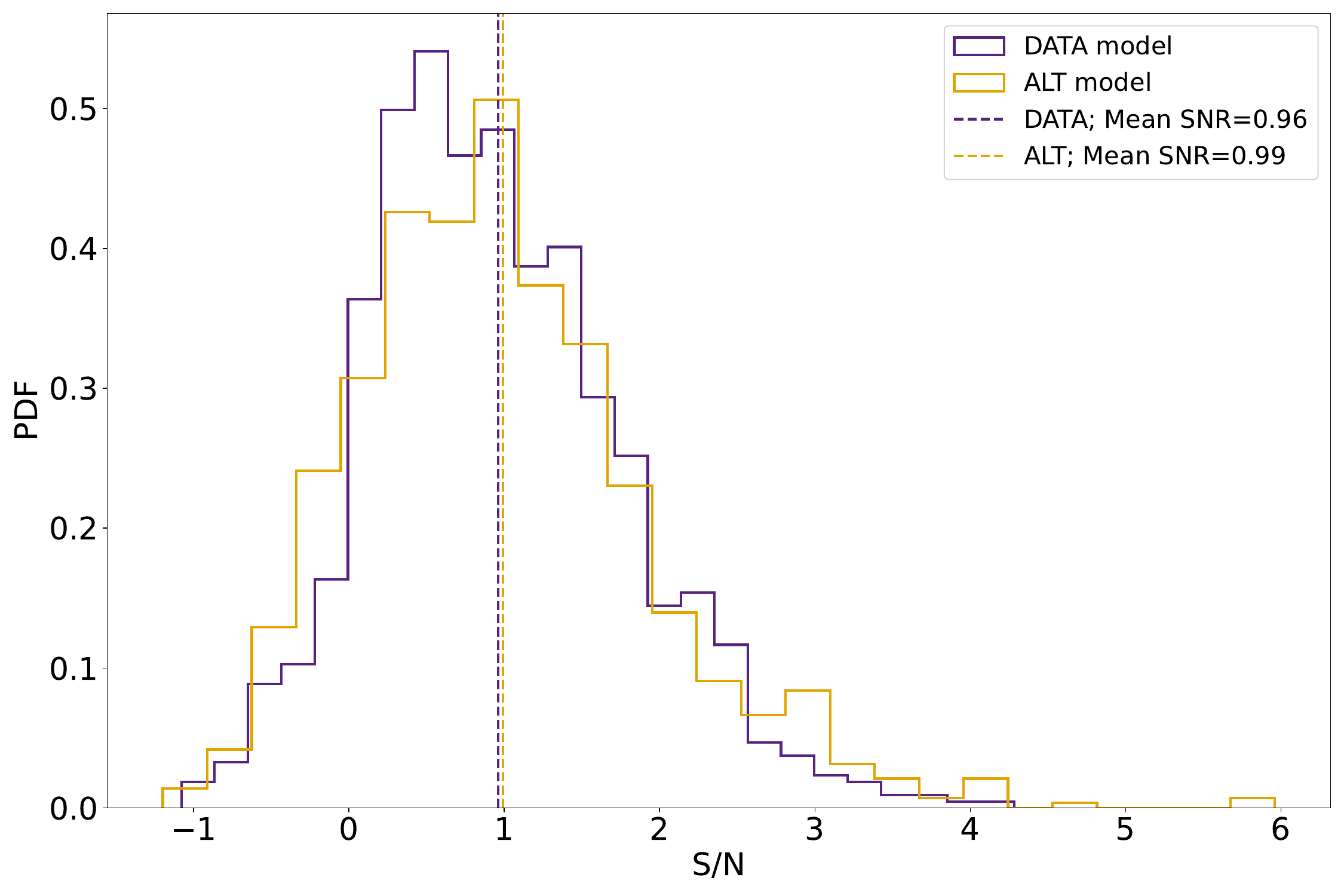}
    \caption[S/N distribution attained from the noise marginalised optimal statistic.]{S/N distribution attained from the noise marginalised optimal statistic. The distribution was calculated from $1000$ samples taken from the Bayesian search for a CURN. The mean S/N attained through this method is considerably lower than that attained using the FP values.}
    \label{fig: noise_marg_SNR}
\end{figure}

We assess the significance of the optimal statistic after noise marginalisation as described in \citet{2018PhRvD..98d4003V}. This method has been shown to obtain a more reliable measurement of the cross-correlated amplitude. To perform this analysis, we use the posterior samples from a Bayesian analysis which models the intrinsic pulsar noise processes in the MPTA, while also searching for a common uncorrelated noise process with a variable spectral index. We perform this assessment on the DATA and ALT models as the ER model did not provide a sufficiently high signal to noise ratio to warrant further frequentist analyses. The distribution of the S/N attained through this marginalisation is shown in Figure \ref{fig: noise_marg_SNR}. In this search, we find a mean S/N and a standard deviation of $0.96$ and $0.81$ for the DATA model, and a corresponding $0.97$ and $0.94$ for the ALT model. We note that while the mean values of the noise marginalised OS S/N are considerably lower than the FP value S/N, both distributions have long tails towards larger S/Ns where the FP solution resides. 

We note that the S/N that is reported through this analysis cannot be confidently used to assess detection probability. Instead, alternate methods are required to empirically derive a significance. We calculate a probability (i.e., a \textit{p}-value), using noise realisations from the pulsar timing data to establish a null space to which the observed correlations can be compared. In standard PTA analyses, a combination of \textit{phase shifts} and \textit{sky scrambling} are employed \citep{2016PhRvD..93j4047C, 2017PhRvD..95d2002T}.  

\begin{figure}
\subfloat
  \centering
  \includegraphics[width=\columnwidth]{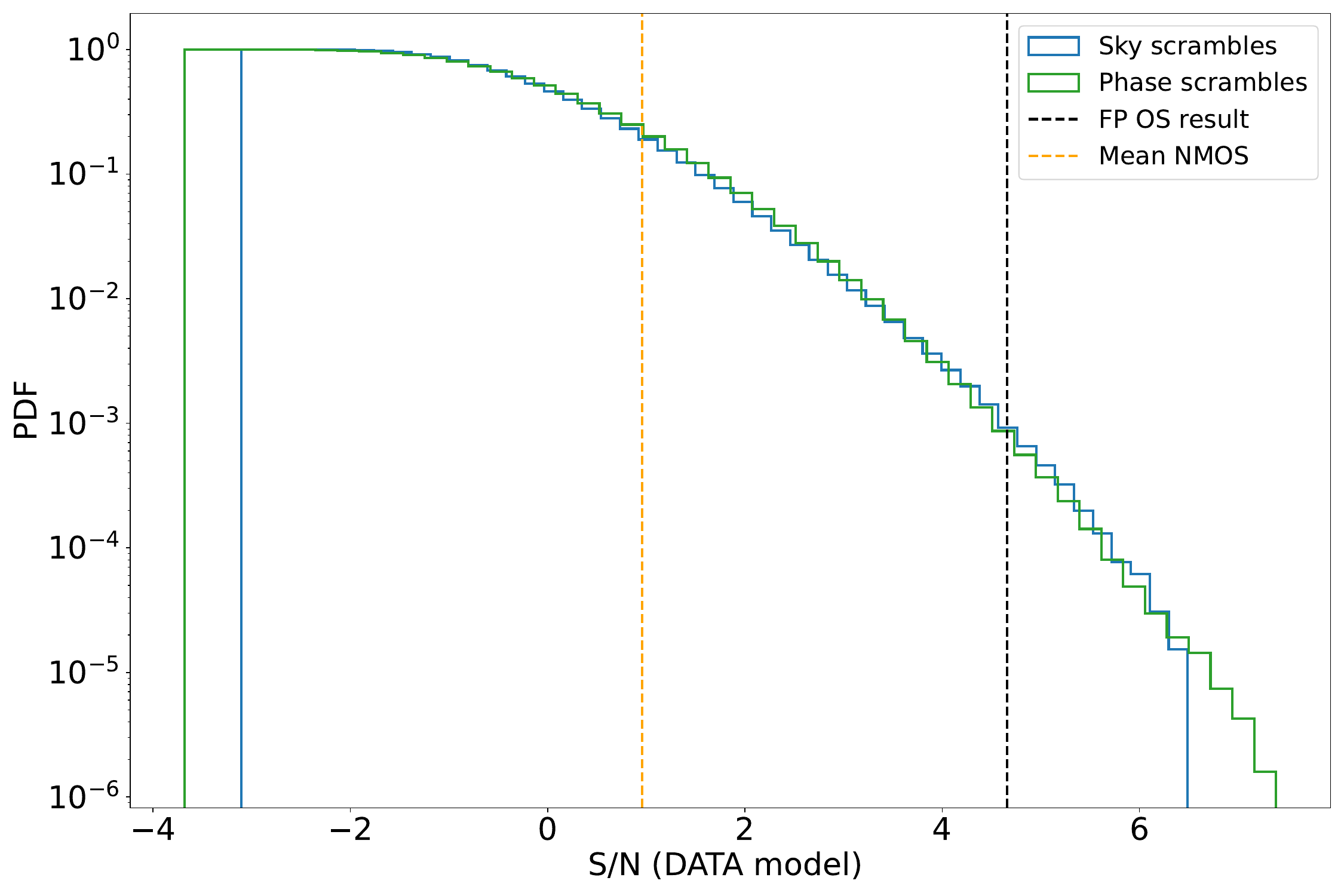}
  \label{fig: DATA_scrambles}

\subfloat
  \centering
  \includegraphics[width=\columnwidth]{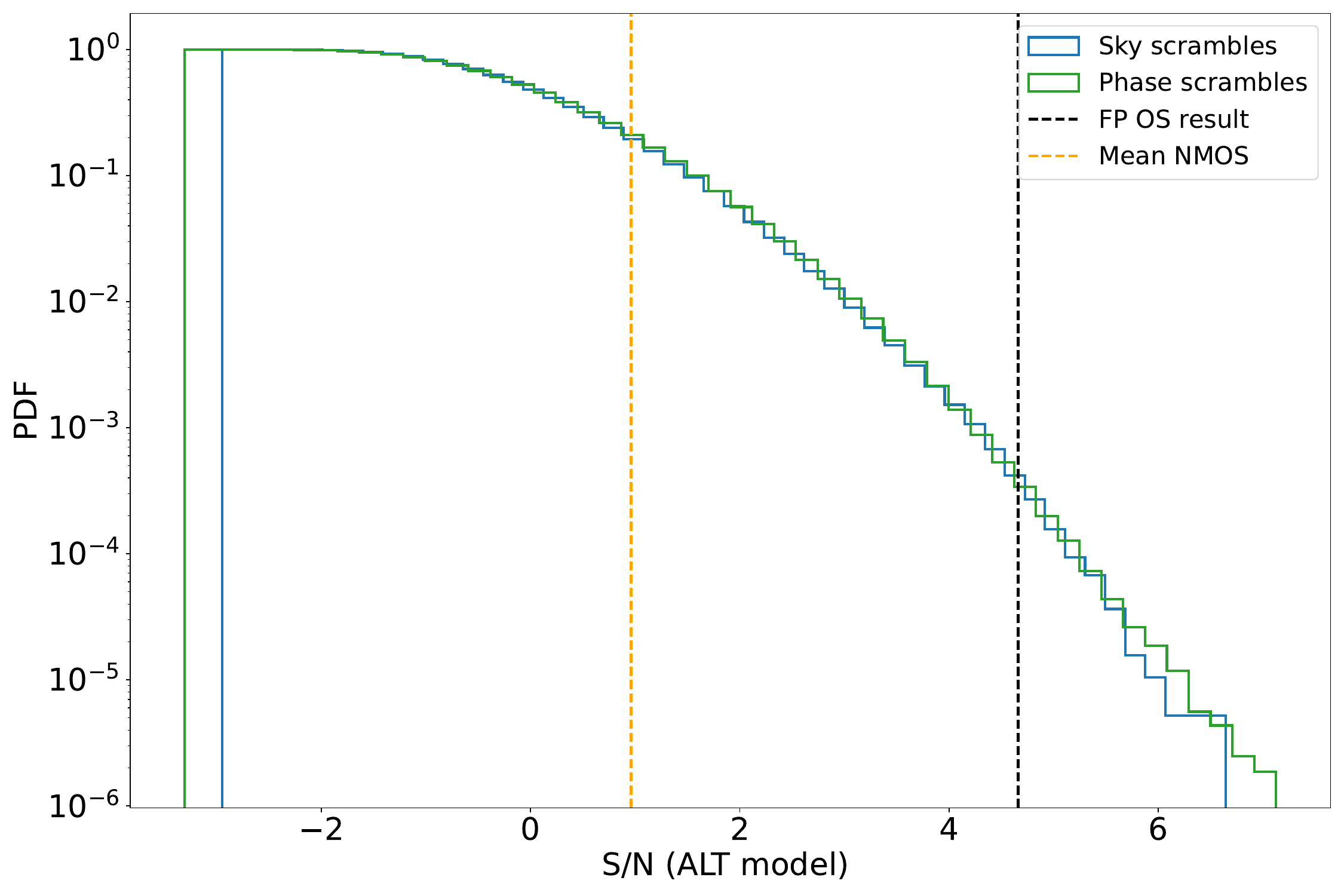}
  \label{fig: ALT_scrambles}

\caption[S/N distribution of sky and phase scrambles used to assess the significance of frequentist estimate of inter-pulsar angular correlations.]{Distribution of the sky (green) and phase (blue) scrambles calculated from the MPTA data, for the DATA (top) and ALT (bottom) assumptions. The black-dashed line corresponds to the S/N calculated from the optimal statistic using the FP values, the orange line corresponds to the mean of the noise marginalised S/N distribution. The mean noise marginalised S/N is considerably less significant than that attained using the FP values.}
\label{fig: ch6_scrambles}
\end{figure}

Assessing the DATA model first, we find that $102$ out of $\sim 1.3\times10^{5}$ randomised pulsar sky positions result in a S/N that is greater than the value we have measured, corresponding to a p-value of approximately $8 \times10^{-4}$. Through phase shifting we found a similar significance at a p-value of $6.4 \times 10^{-4}$ from $1.9 \times 10^{6}$ samples. Assessing the ALT model in a similar manner, we find p-values on the order of $1.4 \times 10^{-3}$ and $1.3 \times 10^{-3}$ through sky and phase scrambling respectively. The distributions of these bootstrapped methods are shown in Figure \ref{fig: ch6_scrambles} and correspond to $\sim 3.2 - 3.4\sigma$, representing evidence for nHz-frequency gravitational waves assuming the aforementioned FP values. While these values are intriguing, they were measured using the FP values derived from Bayesian analyses, and decrease significantly where we instead apply the mean values from the noise marginalisation process. The null distributions displayed in Figure \ref{fig: ch6_scrambles} were also calculated using FP values. Where we assume the median values of the Bayesian analyses to establish the null distributions instead, we find comparable estimates for the significance.

We note that variability between pulsars in achieved timing precision and intrinsic pulsar noise processes reduces the number of \textit{effective} pulsar pairs in a PTA. For the MPTA, taking this into account for the DATA model results in only $608$ effective pulsar pairs in contrast to the total $3403$, and for the ALT model only $482$, impacting the suitability of standard scrambling techniques \citep{2023arXiv230504464D}. To account for this, an additional \textit{super scramble} technique has been developed \citep{2023arXiv230504464D} which allows for a greater number of independent realisations. We have not employed this as subsequent work by \citet{2024arXiv240313175D} suggests that correlated scrambles probably produce conservative values even in the presence of misspecified noise.

\subsection{Bayesian assessment of isotropic angular correlations}
\label{subsection: Bayesian_isotropy}

For both the DATA and the ER models we perform ensemble Bayesian analyses, searching for a common, achromatic signal possessing HD correlations in the MPTA data set. We use the \texttt{PTMCMC} sampler \citep{2009MNRAS.395.1005V, 2013MNRAS.428.1147V, 2014MNRAS.437.3004L, 2014PhRvD..90j4012V, justin_ellis_2017_1037579} to determine the posterior distributions of all time-correlated processes while holding the white noise parameters fixed at the values reported in \citet{MPTA2024_data+noise}. We consider both models to demonstrate the suitability of our noise model selection technique, and to investigate if the DATA model is more representative than the conservative ER model. 
% Additionally, we analyse the ER model to mitigate against misspecification of sub-threshold noise processes into the common MPTA signal, although that may be less likely to occur where the signal must possess angular correlations.
For the DATA model, we estimate an amplitude of $\log_{10}\mathrm{A_{HD}^{DATA}}=-14.28^{+0.23}_{-0.30}$ at a spectral index of $\gamma_\mathrm{{HD}}^\mathrm{DATA}=4.50^{+1.00}_{-0.93}$, consistent with the complementary CURN constraints for this model. The ER model is likewise consistent with the CURN derived measurements of the common signal in the MPTA data, measured at an amplitude of $\log_{10}\mathrm{A_{HD}^{ER}}=-14.31^{+0.21}_{-0.42}$ and spectral index $\gamma_\mathrm{{HD}}^\mathrm{{ER}}=3.40^{+1.45}_{-0.98}$. 

We assess the evidence for HD correlations by comparing the Savage-Dickey Bayes factors that are calculated for Bayesian analyses assessing a CURN signal, and one assessing a common signal possessing HD correlations. If an HD process is supported, it should have a higher Bayes factor, with the ratio of Bayes factors representing the evidence in support for (or against) Hellings-Downs correlations. In neither case do we find support for Hellings-Downs correlations in this MPTA data set. For the ER model we report $\ln\mathcal{B}^\mathrm{HD+CURN}_\mathrm{CURN}=-0.21$. For the DATA model, we find that there is not sufficient posterior support at low amplitudes to calculate a direct Bayes factor via the Savage-Dickey method. We can, however, provide limits on the Bayes factors. Comparing the results in this way, we find that both models possess comparable lower limits to the PTA intrinsic red noise (IRN) at $\ln\mathcal{B}^\mathrm{HD+/CURN}_\mathrm{IRN}<13.4$. As both models share this lower limit, there is a similar ambiguity in assessing which is favoured by the data.

\begin{figure}
    \includegraphics[width=\columnwidth]{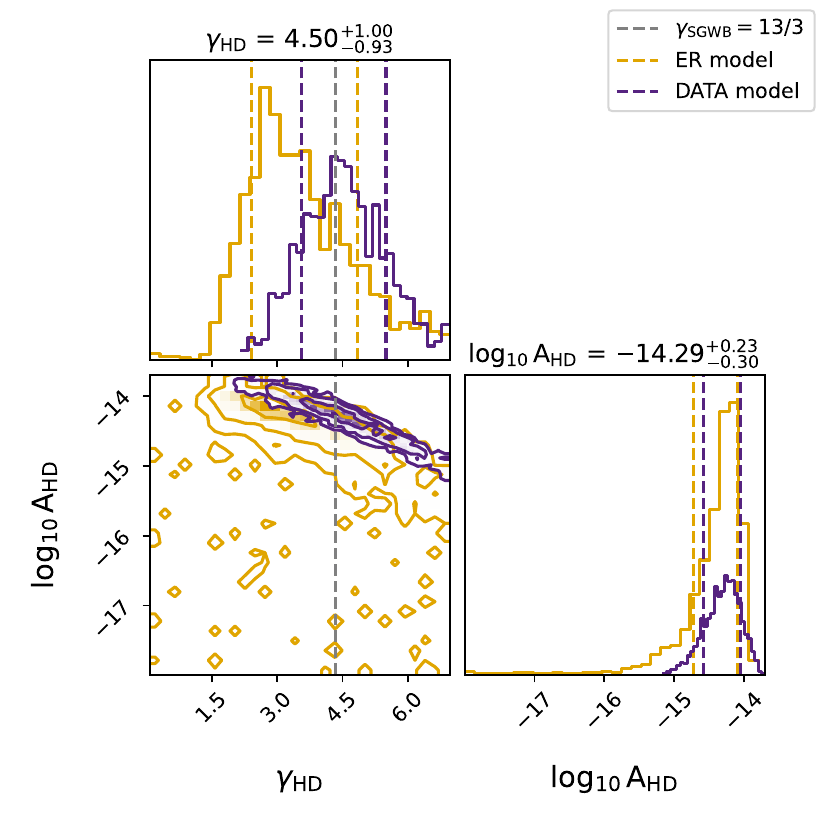}
    \caption[Marginalised posterior distributions of an HD correlated common signal in the MPTA data.]{Marginalised posterior distributions of the logarithm of the strain amplitude ($\log_{10}\mathrm{A_{HD}}$) and spectral index ($\gamma_\mathrm{HD}$) of a Hellings-Downs correlated common signal in the MPTA data, assessed both with the DATA (purple) and ER (gold) models. The contours are the $1\sigma$, $2\sigma$, and $3\sigma$ confidence regions of the posterior distributions, and the values reported above the one-dimensional posteriors are the median and $1\sigma$ values of the DATA model. The spectral indices of both models are consistent with that expected of an SGWB, which we have overlaid for comparison (grey, dashed).}
    \label{fig: HD_Bayesian_posts}
\end{figure}

However, the constraints on the common spectral properties do not appear any more pronounced where considering HD correlations as opposed to not. The posteriors of these processes are shown in Figure \ref{fig: HD_Bayesian_posts}.

\subsection{\textit{Pseudo-Bayesian} assessment of isotropic angular correlations}
\label{subsection: pairwise_isotropy}

The MPTA possesses the largest number of pulsar pairs of any individual PTA, but the shortest observing span over which to constrain stochastic processes. Due to this, the \textit{pseudo-Bayesian} pairwise analysis offers us an opportunity to assess the correlations that may be present in the MPTA in a manner that could be more suitable than the other analyses we present here: one that places a greater emphasis on the correlations between the pairs in the array, rather than on the stochastic processes themselves.

By sampling the shared correlations and amplitudes between each of the $3403$ pulsar pairs, we can factorise the likelihoods of these constraints and redistribute them into bins corresponding to their angular separations on the sky. From this, we are able to calculate a likelihood ratio ($\mathcal{L}^\mathrm{HD}_\mathrm{0}$) from the factorised posteriors to determined if they are more suited to HD correlations than not, following the technique presented in \citet{2023ApJ...951L...6R}.

\begin{figure}
    \includegraphics[width=\columnwidth]{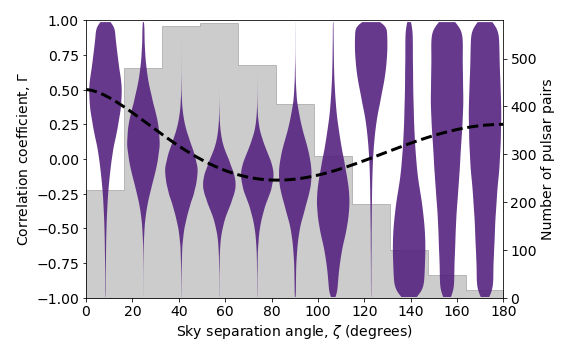}
    \caption[Factorised correlations of each pulsar pair in the MPTA using the ER model.]{Factorised correlations of each pulsar pair in the MPTA under the assumptions of the ER model. Through this technique, we recover marginal support for Hellings-Downs like correlations with the ER model.}
    \label{fig: ER_model_pairwise}
\end{figure}

The number of pulsar pairs in the MPTA may make this analysis more prone to error propagation than in other PTA experiments. While the number of pulsar pairs is likely to make the MPTA sensitive to interpulsar correlations, the process of factorising hundreds of posteriors together may result in otherwise insubstantial uncertainties in the posterior constraints compounding, and result in non-representative factorised correlations. Given this, we only assess the conservative ER model, which exhibits the smallest margin of support for HD correlations in the frequentist analysis. We find marginal support for HD correlations, at $\mathcal{L}^\mathrm{HD}_\mathrm{0} \sim 7$. The factorised angular correlations derived from this technique are found in Figure \ref{fig: ER_model_pairwise}.

\subsection{Searching for a single source}
\label{subsection: ch6_results_single_source}

\begin{figure*}
    \includegraphics[width=\linewidth]{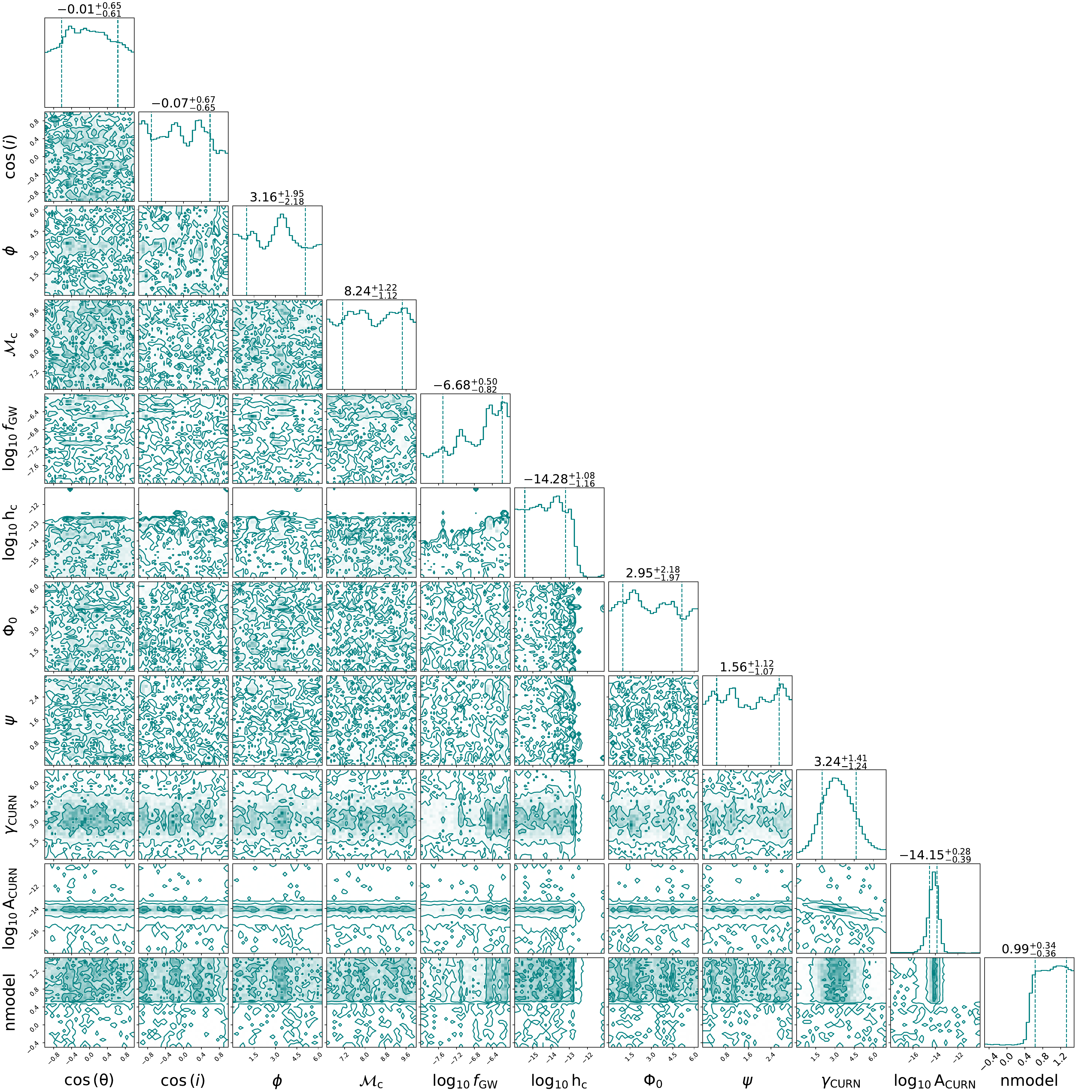}
    \caption[Marginalised posterior distributions of the CW and CURN signal parameters.]{Marginalised posterior distributions of the CW and CURN signal parameters using the ALT model. We utilise the \texttt{hypermodel} framework of the \textsc{ptmcmc} sampler to naively calculate a Bayes factor between the models, represented by the parameter labelled {\em nmodel}. The median and $1\sigma$ values of the CW and CURN parameters are displayed above their one-dimensional posterior distributions. We note that a smoothing factor has been applied to the one-dimensional distributions.}
    \label{fig: ch6_SMBHB_CRN_posteriors}
\end{figure*}

The ambiguity we observe in our results indicate that the signal might not be well modelled by an isotropic background. Given this, we also search for a CW as the progenitor of the common signal. To evaluate this, we have searched this MPTA data set for a single SMBHB, using the methods described in Section \ref{subsection: ch6_single_source_methods}.
%modelling the potential source as an isolated SMBHB. 
In Figure \ref{fig: ch6_SMBHB_CRN_posteriors} we show the marginalised posterior distributions of a Bayesian analysis searching for both a CURN and an isolated SMBHB. By utilising the \texttt{hypermodel} framework of the \texttt{PTMCMC} sampler, we can measure a Bayes factor directly. We find that the data strongly favours a model described by both a CW and a CURN as opposed to only that of a CW, with a Bayes factor of $\ln\mathcal{B}^\mathrm{SMBHB+CURN}_\mathrm{SMBHB} = 9.12$. By taking the Savage Dickey Bayes factor of both signals, we can also calculate the degree to which the data prefers the CURN model as opposed to the SMBHB model. We find this to be $\ln\mathcal{B}^\mathrm{CURN}_\mathrm{SMBHB} = 7.76$.

While the data do not support the presence of an individual source, it is worth noting that some posterior parameters show interesting constraints, particularly the frequency of the gravitational-wave signal, which may indicate a preferred emission frequency in the MPTA data set.

\section{Discussion}
\label{ch6: discussion}

In this analysis, we have investigated the presence of nHz-frequency gravitational waves through the assessment of apparent inter-pulsar correlations well described by the Hellings-Downs overlap reduction function. 

\subsection{Independent assessment of nHz-frequency gravitational waves}

The common uncorrelated noise process recovered using the DATA model is consistent with that which is recovered using the ER model in \citet{MPTA2024_data+noise}. As discussed in \citet{MPTA2024_data+noise}, this is inconsistent with the recovered common noise properties recently published by other PTAs by at least $\sim1.4 \sigma$. It remains unclear if the amplitude measurement of the CURN is physical, or if it is the result of the short MPTA data set as compared to other PTAs. If it is physical, the larger amplitude could be an extension of the amplitude growth observed in the PPTA data \citep{2023ApJ...951L...6R}, and implied in the analysis performed by the EPTA+InPTA \citep{2023arXiv230616214A}. If the amplitude growth is of a physical origin, the relative temporal independence of the data used in this analysis would explain why we have measured a larger amplitude.  

The analysis in Section \ref{ch6: results} presents an assessment of nHz-frequency gravitational waves with a data set that is largely temporally independent from other searches that have been performed with the techniques utilised in this work. Using the DATA model, the inter-pulsar correlations recovered for the $83$ pulsars imply a dimensionless strain amplitude of $\mathrm{h_{c}} = 7.5 \times 10^{-15}$, with a reported S/N of $4.6$ when fixing the noise models at the FP values. Through the use of sky scrambling and phase shifting methods, we calculate the false alarm probabilities of this measurement to be $p= 8 \times 10^{-4} \ \mathrm{and} \ 6.4 \times 10^{-4}$ respectively.

However, through a noise-marginalised approach, where the covariance of the pulsar intrinsic noise processes are taken into account in the calculation of the optimal statistic, we find that the mean S/N for both the DATA and ER models is $\sim 1$. This is considerably lower than the S/N reported using the FP values, and implies that the convergence on the spectral properties of some MPTA noise processes may be too broad to be well-described by the FP solution. The similarly broad global constraints of the common signal may also play a part in reducing the mean S/N, as this describes a large parameter space that the several hundred free parameters we consider could find support in.

The short time span that the MPTA possesses results in this signal only being significant in three frequency bins, and the analysis of the signal using the frequentist approach is limited to these. The measurement from the optimal statistic analysis, where the FP spectral index is employed, is equivalent to an amplitude of $\log_{10}\mathrm{A_{DATA}} = -14.12^{+0.04}_{-0.05}$, which is consistent within uncertainties of the DATA model measurement of the CURN, $\log_{10}\mathrm{A_{CURN}^{DATA}} = -14.25^{+0.21}_{-0.34}$. Assuming that the FP value solution is correct, we conclude that the CURN is dominated by the GW signal and not by misspecified pulsar noise \citep{2022MNRAS.516..410Z}.

Assuming a spectral index of $13/3$, the value associated with an isotropic SGWB formed of inspiralling SMBHBs, the optimal statistic amplitude is measured at $\log_{10}\mathrm{A_{DATA}} = -14.32^{+0.07}_{-0.08}$, slightly smaller than the FP solution due to the modestly steeper spectral index. Further, the Bayesian analysis of HD correlations in the MPTA data set reports a similar amplitude of the common signal, at $\log_{10}\mathrm{A_{HD}^{DATA}} = -14.28^{+0.23}_{-0.30}$. The consistency between these measurements, although the Bayes factor between them remains inconclusive, is supportive that the signal could be of a GW origin. 

During this analysis, the \textit{pseudo-Bayesian} pairwise technique was also employed. Using our most conservative model, we assessed if an analysis stream that is more dependent on the large number of pulsar pairs in the MPTA would be more sensitive to a correlated signal in the data than other techniques. Evidence for the presence of HD correlations in the data was found at $\mathcal{L}^\mathrm{HD}_{0}\sim7$, which could be interpreted as marginal support for the correlations where the Bayesian assessment in Section \ref{subsection: Bayesian_isotropy} had none. However, there is no clear indication that the correlations returned by this technique are more pronounced, or better constrained, than the frequentist FP value solution for this model.

The temporal independence of the MPTA data set potentially provides further support to the presence of detectable gravitational radiation in the nHz-frequency band. Naively it may be expected that MPTA pulsars, and therefore our data set, are also relatively independent as many are at Southern declinations as opposed to the largely Northern declination pulsars observed by most other PTAs. However, the most precisely timed pulsars will influence these results the most, and many of these (such as PSR~J1909$-$3744) are observed by several PTAs. 

For an isotropic background of GWs emitted by SMBHBs, the amplitude that we have measured is thought to be largely a function of the masses of the SMBHBs that make up the ensemble distribution \citep{2013MNRAS.433L...1S, 2014ApJ...789..156M}. If this is indeed the source of the signal we observe, the amplitude that we have measured is within the $95\%$ confidence limit of models presented in \citet{2013MNRAS.433L...1S}, when we consider the lowest frequency resolved by the MPTA data set. However, recent work by \citet{2023arXiv230600898S} has presented that, in simulations where SMBH seeds are able to establish their mass early, the predicted amplitude of the background may be larger by as much as a factor of three. Given this, it may not be unexpected to measure the amplitude of the SGWB that we observe.

\subsection{Under the assumption of a single source}

The ambiguity of the analyses in this data set led us to consider if a single source emitting CWs could explain the apparent angular correlations. A search was performed on the MPTA data for a circularly inspiralling SMBHB emitting continuous gravitational waves; however, there was a clear preference for a model that not only contained a continuous waveform, but one that also included a common stochastic signal in the form of a CURN. The marginalised posterior distribution of this model in Figure \ref{fig: ch6_SMBHB_CRN_posteriors} shows clear preference for the CURN$+$CW model, and the posterior distribution is unconstrained for most of the CW model parameters. Of note, however, is the marginal support for a CW at higher frequencies shown in the one-dimensional posterior distribution of the gravitational wave frequency ($\log_{10}f_\mathrm{GW})$. This constraint also appears to be covariant with the sky-location CW parameter distributions, certainly motivating a more thorough assessment of alternate explanations of a correlated signal other than a isotropic gravitational wave background (see \citet{MPTA2024_aniso}).

\subsection{Comparison to other PTAs}

In our search for a gravitational wave background using the optimal statistic, we find the characteristic strain amplitude in the MPTA data is $\log_{10}\mathrm{A_{DATA}} = -14.12^{+0.04}_{-0.05}$ assuming FP values of the common signal in the data. In \citet{MPTA2024_data+noise}, we discussed that fixing the spectral index of the common uncorrelated process at the nominal value of $13/3$, associated with an isotropic SGWB formed from an ensemble of inspiralling SMBHBs, resulted in a minimum deviation of $1.35 \,\sigma$ as compared to other PTA experiments.

Assessing the optimal statistic amplitude under these assumptions, we measure a characteristic strain log-amplitude of $\log_{10}\mathrm{A_{DATA}} = -14.32^{+0.07}_{-0.08}$, consistent with our previous result within $1\sigma$ uncertainties. However, we acknowledge the apparent sensitivity of the OS amplitude to the parameter values and models that are chosen, which can be observed between Figures \ref{fig: HD_all_sky_DATA} and \ref{fig: OS_ER_ALT}. The Bayesian assessments of a common signal with HD correlations do not appear to share this dependence between our choice of models (Figure \ref{fig: HD_Bayesian_posts}), and so we compare this instead to the results of other PTAs. Isolating the Bayesian search of the DATA model to a small window ($0.1$) around $\gamma=4.333$, we measure a log-amplitude of $\log_{10}\mathrm{A_{DATA}} = -14.39^{+0.11}_{-0.11}$, consistent with our OS measurement but centered at a marginally lower value.

The EPTA+InPTA collaboration measured the amplitude of the signal in their data at a fixed spectral index using the optimal statistic for several versions of their data set. Of these measurements, the largest is $\log_{10}\mathrm{A} = -14.48^{+0.08}_{-0.11}$ \citep{2023arXiv230616214A}, for a data set including only their most recent 9\,yr of data. The PPTA did not use an optimal statistic analysis in their most recent search, so we compare to the strain amplitude measured in their pairwise search for an angularly correlated signal, which they report to be $\log_{10}\mathrm{A} = -14.69^{+0.05}_{-0.05}$ \citep{2023ApJ...951L...6R}. While NANOGrav used an optimal statistic approach, they did not report a strain amplitude from this assessment. Instead, they report a strain amplitude from their Bayesian analysis of $\log_{10}\mathrm{A} = -14.62^{+0.11}_{-0.12}$ \citep{2023ApJ...951L...8A}. The MPTA measurement is larger in amplitude than all values reported by other PTAs. In comparison to the PTA that reported the next largest amplitude, the EPTA+InPTA, our measured amplitude overlaps within reported uncertainties. However, in the case of the PPTA, which reported the smallest amplitude, we are at a far greater $2.5 \sigma$ discrepancy.

While the differences between the amplitude we present here and that taken by the PPTA is the largest, the PPTA also reported an apparent growth in the strain amplitude of the common signal found in their data. This was identified by splitting their entire dataset into overlapping $9$ and $6$ year intervals, which they referred to as slices. The strain amplitude in their most recent $6$ year time slice was measured to be $\log_{10}\mathrm{A} = -14.57^{+0.11}_{-0.13}$, just $1.2 \sigma$ away from the MPTA measurement. This could suggest that shorter data sets may result in signal measurements with larger amplitudes, however, there is a relative growth in the signal amplitude through the equal slices of data reported in \citet{2023ApJ...951L...6R}, implying a temporal dependence which may be affecting the MPTA measurements. 

The pulsar timing array data analysis with the most temporal overlap with the MPTA is the Chinese Pulsar Timing Array DR1 \citep{2023RAA....23g5024X}. While they provide a constraint on the CURN in their analysis (which is discussed in \citet{MPTA2024_data+noise}), they use an alternate statistic to assess the presence of cross correlated signals that does not provide an estimate of the characteristic strain amplitude. It is not possible to directly compare our results to theirs.

\subsection{Interpreting the MPTA results}

Using methods that are standard practice in PTA experiments, we have reported evidence of a correlated signal in a $4.5$-year MPTA data set, with inter-pulsar correlations indicative of a gravitational wave background. The signal is found using data-informed models describing astrophysically motivated noise processes, is visually apparent across the angular separation range, and has a significance equivalent to $\sim3.4 \sigma$. When using a more conservative model, we find a comparable significance of $\sim3.2 \sigma$. An assessment of the signal under both of these model assumptions results in visually apparent HD correlations when using FP values, providing confidence that the signal may be of GW origin. However, assessing the most conservative model that we consider (ER) does not result in apparent correlations.

While this result is likely to be statistically robust, especially as the background-estimation techniques used to derive the significance are thought to be conservative \citep{2024arXiv240313175D}, other aspects of our analysis do not show the same support for the signal in the data. Marginalising over correlated noise constraints diminishes the significance of the signal, and the results from Bayesian ensemble searches for a common signal possessing HD correlations are similarly insignificant. The difference in significance between our results is notable, even if the constraints on the amplitude measurements are consistent with each other. 

The cause of these differences is not well understood. It may be that the short data span of the MPTA places a heavy toll on our GW searches and noise analysis. Each of the methods we employ here is predicated on some form of Bayesian analysis that requires the ability to precisely and accurately constrain the noise processes. While the sensitivity to these processes should increase quickly for the MPTA given the comparatively greater timing precision than other PTA experiments, the processes still require several significant frequencies to be confidently constrained. For many pulsars in the MPTA data set, this is likely not yet achieved, leading to broad posterior constraints. There is also considerable covariance between different families of noise models (e.g., dispersion measure noise and other chromatic noise with some alternate chromaticity), which can lead to further confusion in the analysis.

The broad constraints on the noise processes result in many possible combinations of parameter values that well model both the intrinsic pulsar noise and any common noise process across the array. In many of these combinations, the parameter values will differ from our chosen values, some moreso than others. In these cases, it can be expected that the description of the intrinsic pulsar noise does not provide probabilistic support for a Hellings-Downs correlated signal. Naturally, a PTA that has a greater number of noise parameters that are prone to weaker constraints, such as the MPTA, will be more susceptible to this.

The variability of our results with different noise models highlights a promising area for future study. For future work, we envision a single, unified noise model with sufficient flexibility to describe the noise in MPTA pulsars without artificially inflating error bars with overly conservative assumptions. Such a model could employ hierarchical modelling to model the population properties of pulsars \citep{2022ApJ...932L..22G, 2024arXiv240605081V}. It could marginalise over multiple sub-models, some of which include additional noise processes, e.g., achromatic red noise. In this way, we can use the data to help determine the model complexity. The ambiguity between model selection, and the desire for conservatism when constructing said models, may mean the correct approach is to marginalise over all models we consider. The strategy we have employed in the design of our models is perhaps a crude analogue to this, which could naturally be developed to involve all unmodelled processes across a full PTA analysis. We leave this for future work.

While the specific issues with this analysis are not well understood, it should be possible to resolve the nature of any signal present in the MPTA data set soon.
Firstly, the ability to discriminate signals should increase rapidly as MPTA observations continue. The strength of red noise processes (both a gravitational-wave background and the processes most covariant with it) increases quickly with time, relative to white noise processes associated with ToA measurement uncertainty and pulse jitter noise \cite[][]{2013MNRAS.428.1147V}. Using the methodology outlined in \citet{2013CQGra..30v4015S}, \citet{spiewak2022} demonstrated that for a background comparable in amplitude to that which we measure here, the MPTA is currently well described by the moderate or weak signal regimes. As such, even modest increases in the length of MPTA data sets are fractionally large compared to the other, more established, PTA experiments.

Secondly, it is possible that the results presented here motivate the need for new pulsar timing analysis methods, that are better able to assess the presence of a stochastic background in the presence of a cacophony of noise processes.
While power-law Gaussian processes are commonly used to model stochastic processes in pulsar timing analysis, it is unclear if they are appropriate for both achromatic and chromatic noise \cite[e.g.,][]{1985ApJS...59..343C,2015ApJ...808..113C}. We see evidence for this through the identification of Gaussian and annual variations in some of the pulsars in our sample. Similarly, population level inference of noise properties might be able to reduce covariance between intrinsic spin noise, chromatic noise, and any gravitational wave signal \cite[][]{2024arXiv240313175D}.

Lastly, the nature of the signals will benefit through comparison and combination with other PTA data sets. Many of the best timed pulsars in the MPTA are observed by other PTA experiments. The combination of the data analysed here and presented in \citet{MPTA2024_data+noise} in the planned IPTA third data release will greatly assist the efforts of all collaborations in determining the nature of the various stochastic signals.

\section{Conclusions}
\label{ch6: conclusions}

In this analysis, we have presented an assessment of the presence of nanohertz-frequency gravitational waves in the MeerKAT Pulsar Timing Array, using a data set largely independent from other PTA experiments, sensitive at different GW frequencies. Using the data from $83$ pulsars in the MPTA, observations of which were collected over a period of $4.5$ years from the MeerKAT radio telescope, we find an angularly correlated signal consistent with the expected Hellings-Downs correlations. Using our most significant model, the amplitude of the cross-correlated signal is estimated to be $\mathrm{A_{GW}} = 7.5^{+0.8}_{-0.9} \times 10^{-15}$ measured at the FP spectral index of the process, and $\mathrm{A_{GW}} = 4.8^{+0.8}_{-0.9} \times 10^{-15}$ where measured at the predicted spectral index of a SMBHB dominated SGWB, $\gamma=13/3$. The S/N measured at FP values for the presence of this signal in the data is $4.6$, with false alarm probabilities corresponding to $\mathrm{p}=8\times10^{-4}$ and $\mathrm{p}=6.4\times10^{-4}$, calculated using the background estimation sky and phase scrambling techniques, respectively. We also undertake a search for a single radiating source in the form of a loud SMBHB, however, we find it is not supported compared to a model including a common uncorrelated signal. The false alarm probabilities that we report under the assumption of an isotropic background correspond to a significance of $\sim3.4 \sigma$, although it is important to note that these results are particularly sensitive to assumptions made in the construction of the intrinsic pulsar noise models. Further, the application of practices such as noise marginalisation and Bayesian ensemble analyses used in the assessment of the MPTA results greatly diminishes their significance, implying that the data are ambiguous to the presence of Hellings-Downs correlations. Given these caveats, interpreting the nature of the apparent correlated signal in the MPTA data is non-trivial. Further work using greater amounts of MPTA data, and in combination with IPTA data, is required to build confidence in our results.
Nevertheless, our results demonstrate the ability of sensitive, high-cadence PTA experiments to broaden the nanohertz-frequency gravitational wave window, and enable access to new GW sources and GW source classes, while also increasing the sensitivity of international efforts.

\section*{Acknowledgements}

The MeerKAT telescope is operated by the South African Radio Astronomy Observatory (SARAO), which is a facility of the National Research Foundation, an agency of the Department of Science and Innovation. SARAO acknowledges the ongoing advice and calibration of GPS systems by the National Metrology Institute of South Africa (NMISA) and the time space reference systems department department of the Paris Observatory.
MeerTime data is stored and processed on the OzStar and Ngarrgu Tindebeek supercomputers, operated by the Swinburne University of Technology.
This work was undertaken as part of the Australian Research Council Centre of Excellence for Gravitational Wave Discovery (CE170100004 and CE230100016).  RMS acknowledges support through ARC Future Fellowship FT190100155.
GT acknowledges financial support from “Programme National de Cosmologie and Galaxies” (PNCG), “Programme National Hautes Energies” (PNHE) and "Programme National Gravitation, Références, Astronomie, Métrologie" funded by CNRS/INSU-IN2P3-INP, CEA and CNES, France. We acknowledge financial support from Agence Nationale de la Recherche (ANR-18-CE31-0015), France.
MK acknowledges support by the MPG and the CAS-MPG Legacy Programme. Part of the PTUSE hardware was provided by MPIfR.
KG acknowledges continuing valuable support from the Max-Planck society.
KG acknowledges the support from the International Max Planck Research School (IMPRS) for Astronomy and Astrophysics at the Universities of Bonn and Cologne.
V.V.K acknowledges financial support from the European Research Council (ERC) starting grant ""COMPACT"" (Grant agreement number 101078094).
FA acknowledges that part of the research activities described in this paper were carried out with the contribution of the NextGenerationEU funds within the National Recovery and Resilience Plan (PNRR), Mission 4 - Education and Research, Component 2 - From Research to Business (M4C2), Investment Line 3.1 - Strengthening and creation of Research Infrastructures, Project IR0000034 – “STILES -Strengthening the Italian Leadership in ELT and SKA”.
Pulsar research at Jodrell Bank Centre for Astrophysics is supported by an STFC Consolidated Grant (ST/T000414/1; ST/X001229/1).
JS acknowledges funding from the South African Research Chairs Initiative of the Depart of Science and Technology and the National Research Foundation of South Africa. 
PG acknowledges support through SUT stipend SUPRA.
AP acknowledges financial support from the European Research Council (ERC) starting grant 'GIGA' (grant agreement number: 101116134) and through the NWO-I Veni fellowship. 
Funding was provided for the PTUSE machines by INAF-Osservatorio Astronomico di Cagliari under the Grant "iPeska" (P.I. Possenti A.).
Funding was provided for the PTUSE machines by the Max-Planck-Institut f\"{u}r Radioastronomie (MPlfR), also supported by the MPG-CAS LEGACY programme, Swinburne University of Technology, and the Australian SKA office.
This project utilises the MeerTime data portal, which is supported by Nick Swainston and the ADACS team.
We acknowledge and pay respects to the Elders and Traditional Owners of the land on which the Australian institutions stand, the Bunurong and Wurundjeri Peoples of the Kulin Nation.
We thank Rutger van Haasteren, Steve Taylor, Siyuan Chen and Levi Schult for useful comments and discussions.

%%%%%%%%%%%%%%%%%%%%%%%%%%%%%%%%%%%%%%%%%%%%%%%%%%
\section*{Data Availability}

All data used in this work is available courtesy of AAO Data Central (\url{https://datacentral.org.au/}) at \url{https://doi.org/10.57891/j0vh-5g31}. 
The data provided includes sub-banded ToAs, the full data archives used to construct this data release, and the ephemerides that have been used to perform timing. Also included are the frequency resolved portraits used to calculate the ToAs used for this work.

The archives and portraits are in \textsc{psrfits} file format. The ephemerides are in a standard \texttt{ascii} text file format, and the arrival times are supplied as IFF data. 



\bibliographystyle{mnras}
\bibliography{ref} % if your bibtex file is called example.bib

%\begin{thebibliography}{99}
%\end{thebibliography}

%%%%%%%%%%%%%%%%%%%%%%%%%%%%%%%%%%%%%%%%%%%%%%%%%%

%%%%%%%%%%%%%%%%% APPENDICES %%%%%%%%%%%%%%%%%%%%%

%\appendix



%%%%%%%%%%%%%%%%%%%%%%%%%%%%%%%%%%%%%%%%%%%%%%%%%%

% Don't change these lines
\bsp	% typesetting comment
\label{lastpage}
\end{document}